\documentclass[conference]{IEEEtran}
\IEEEoverridecommandlockouts
\usepackage{cite}
\usepackage{amsmath,amssymb,amsfonts}
\usepackage{bm}
\usepackage{algorithmic}
\usepackage{graphicx}
\usepackage{textcomp}
\usepackage{subcaption}
\usepackage{xcolor}
\usepackage{stfloats}
\usepackage{balance}
\renewcommand\a{\bm{a}}%
\renewcommand\d{\bm{d}}%

\newcommand\f{\bm{f}}

\newcommand\n{\bm{n}}

\newcommand\w{\bm{w}}

\newcommand\z{\bm{z}}

\newcommand\lambdab{\boldsymbol{\lambda}}

\newcommand\A{\bm{A}}

\renewcommand\H{\bm{H}}%

\newcommand\Z{\bm{Z}}

\newcommand\Xt{\bm{\mathcal{X}}}

\newcommand\THETA{\bm{\Theta}}

\newcommand\CC{\mathbb{C}}
\newcommand\RR{\mathbb{R}}

\newcommand\sysF{\mathcal{F}}

\newcommand\sysN{\mathcal{N}}

\definecolor{JungleGreen}{rgb}{0,0.7,0.4}
\definecolor{deepPurple}{rgb}{0.5,0.1,0.5}

\newcommand{\revTwo}[1]{\textcolor{black}{#1}}

\newcommand\kron{\mathop{\otimes}}         

\renewcommand\max[1]{\underset{#1}{\mathrm{max}}\,}

\newcommand\argmax[1]{\underset{#1}{\arg\,\mathrm{max}}\,}

\newcommand\rmx{\mathrm{x}}
\newcommand\rmy{\mathrm{y}}
\newcommand\rmf{\mathrm{f}}
\newcommand\rmw{\mathrm{w}}

\def\BibTeX{{\rm B\kern-.05em{\sc i\kern-.025em b}\kern-.08em
    T\kern-.1667em\lower.7ex\hbox{E}\kern-.125emX}}
\begin{document}

\title{Probabilistic Position-Aided Beam Selection for mmWave MIMO Systems\\
\thanks{The authors gratefully acknowledge the support of the German Research Foundation (DFG) under the PROMETHEUS project (reference no. HA 2239/16-1, project no. 462458843).}
}

\author{
	\IEEEauthorblockN{\emph{Joseph K.~Chege$^1$, Arie Yeredor$^2$, and Martin Haardt$^1$}
	}
	\IEEEauthorblockA{$^1$Communications Research Laboratory, Ilmenau University of Technology, Ilmenau, Germany
	}
	\IEEEauthorblockA{$^2$School of Electrical Engineering, Tel Aviv University, Tel Aviv, Israel
	}
    \IEEEauthorblockA{Email: \{joseph.chege, martin.haardt\}@tu-ilmenau.de, ariey@tauex.tau.ac.il
	}
}

\maketitle

\begin{abstract} 
Millimeter-wave (mmWave) MIMO systems rely on highly directional beamforming to overcome severe path loss and ensure robust communication links. However, selecting the optimal beam pair efficiently remains a challenge due to the large search space and the overhead of conventional methods. This paper proposes a \revTwo{probabilistic position-aided} beam selection approach that exploits the statistical dependence between user equipment (UE) positions and optimal beam indices. We model the underlying joint probability mass function (PMF) of the positions and the beam indices as a low-rank tensor and estimate its parameters from training data using Bayesian inference. The estimated model is then used to predict the best (or a list of the top) beam pair indices for new UE positions. The proposed method is evaluated using data generated from a state-of-the-art ray tracing simulator and compared with neural network-based and fingerprinting approaches. The results show that our approach achieves a high data rate with fewer training samples and a significantly reduced beam search space. These advantages render it a promising solution for practical mmWave MIMO deployments, reducing the beam search overhead while maintaining a reliable connectivity.
\end{abstract}


\section{Introduction}

Millimeter-wave (mmWave) communications are a key enabler of next-generation wireless networks, offering high data rates and improved spectral efficiency. 
However, mmWave signals suffer from high path loss and sensitivity to blockages, necessitating the use of highly directional beamforming to establish reliable communication links \cite{rappaport2013millimeter, hong_role_2021}. 
In multiple-input multiple-output (MIMO) systems, selecting the optimal beam pair between the base station (BS) and the user equipment (UE) is critical for maximizing the signal strength and the system performance. 
For initial beam establishment at the UE, traditional methods include exhaustive and hierarchical beam search.
Exhaustive beam search, where the UE measures the quality of all BS beams by sweeping all of its receive beams, suffers from a high overhead and latency, whereas hierarchical beam search \cite{xiao2016hierarchical} reduces the overhead incurred by an exhaustive search, but experiences a degradation in beamforming gain due to its multilevel search \cite{khan2023machine}.

Due to the directional nature of narrow beams and the dependency on line-of-sight (LOS) propagation, knowledge of UE positions may be exploited to reduce the training overhead in mmWave communications \cite{morais2023position}. For example, an inverse fingerprinting method that uses past received power characteristics at UE locations has been shown to reduce beam training overhead in vehicular communications \cite{va_inverse_2018}. Machine learning (ML) approaches based on neural networks (NNs) have also been used in this context to learn the complex mapping between UE positions and optimal beam indices~\cite{ heng_machine_2021}. \revTwo{Additional context information such as UE antenna array orientation \cite{rezaie_deep_2022}, vehicle size \cite{wang2018mmWave}, and traffic density \cite{satyanarayana2019deep} has been exploited in ML-based beam selection approaches.}

In this paper, we propose a novel and efficient \revTwo{position-aided} beam selection method that relies on the statistical relationship between UE positions and optimal beam indices, considering them as random variables drawn from an underlying joint probability mass function (PMF). 
We model the joint PMF as a low-rank tensor via the canonical polyadic decomposition (CPD), which has been shown to admit a na\"{i}ve Bayes interpretation in the context of PMF tensor estimation.
This significantly reduces the number of parameters, enabl\revTwo{ing} a reliable estimate to be obtained using relatively few training samples\revTwo{.} \revTwo{In addition, the parameters of our model consist of probability distributions and are therefore interpretable, in contrast with NN weights and biases}.
We evaluate the proposed method using realistic data generated from Sionna~\cite{hoydis2022Sionna}, a state-of-the-art ray tracing simulator. 
\revTwo{Numerical results} demonstrate that our approach significantly reduces the search space while maintaining a higher data rate. 
%
 
%
%
%

\section{System Model} \label{sec:II}
We consider a mmWave MIMO communication system that operates on the downlink.
The communication scenario consists of a fixed BS and a mobile UE, each equipped with a uniform planar array (UPA).
The UPAs are placed on the $\rmx$-$\rmy$ plane and consist of elements $\{M_{{\rm T}_\rmx}, M_{{\rm T}_\rmy}\}$ for the BS and $\{M_{{\rm R}_\rmx}, M_{{\rm R}_\rmy}\}$ for the UE.
The total number of antenna elements at the BS and UE is $M_{\rm T} = M_{{\rm T}_\rmx} M_{{\rm T}_\rmy}$ and $M_{\rm R} = M_{{\rm R}_\rmx} M_{{\rm R}_\rmy}$, respectively.
The received signal at the UE is given by
\begin{equation}
    y_{\rm R} = \sqrt{P_{\rm T}}\w^\mathsf{H}\H\f s + \w^\mathsf{H} \n,
\end{equation}
where $(\cdot)^\mathsf{H}$ denotes the conjugate transpose, $\f \in \CC^{M_{\rm T}}$ and $\w \in \CC^{M_{\rm R}}$ denote the precoding and combining vectors at the BS and the UE, respectively. 
Furthermore, $P_{\rm T}$ is the transmit power, $\H \in \CC^{M_{\rm R} \times M_{\rm T}}$ is the matrix of channel coefficients, $s \in \CC$ is a pilot symbol with unit power, and $\n \in \CC^{M_{\rm R}}$ denotes a zero-mean circularly symmetric complex Gaussian noise vector with variance $\sigma_n^2$.

\subsection{Channel Model}
We generate the channel coefficients $\H$ using Sionna \cite{hoydis2022Sionna}, an open-source software for simulating the physical layer of wireless and optical communication systems.
In particular, we employ the Sionna ray tracing (RT) module to simulate physically accurate channel realizations for various UE positions within a given propagation environment.
Fig.\,\ref{fig:cov_map} shows the simulation environment (also referred to as the ``scene"), which is the area around the Frauenkirche in Munich, Germany.
The scene contains objects made up of different materials such as concrete, brick, wood, glass, metal, among others, resulting in a rich and complex propagation environment.
The ray tracing tool simulates a number of rays, each corresponding to the various propagation paths between the BS and each UE position.
Each path contains associated information such as the angle of departure (AoD), angle of arrival (AoA), the complex path gain, and the delay.
The channel matrix is constructed by applying a narrowband geometric channel model, i.e.,
\begin{equation} \label{eq: channel}
    \H = \sum_{\ell=1}^{L} \alpha_{\ell} \a_{\rm R}(\theta_{\mathrm{R}, \ell},\phi_{\mathrm{R},\ell}) \a_{\rm T}^\mathsf{H}(\theta_{\mathrm{T}, \ell},\phi_{\mathrm{T},\ell}), 
\end{equation}
where $L$ denotes the number of paths, $\alpha_{\ell}$ the complex path gain, $\theta_{\mathrm{R}, \ell}$ and $\theta_{\mathrm{T}, \ell}$ the elevation AoA and AoD, and $\phi_{\mathrm{R},\ell}$ and $\phi_{\mathrm{T},\ell}$ the azimuth AoA and AoD for the $\ell$-th path. 
We consider $L = 25$ dominant paths consisting of one line-of-sight and $L-1$ non-line-of-sight paths arising from reflection and diffraction.
In addition, $\a_{\rm R}(\cdot)$ and $\a_{\rm T}(\cdot)$ are the receive and transmit array steering vectors, respectively.
Let us define the spatial frequencies in the $\rmx$- and $\rmy$-directions as $\mu^{(i)}_{\rmx} = k\Delta_\rmx \sin(\theta_i)\cos(\phi_i)$ and $\mu^{(i)}_{\rmy} = k\Delta_\rmy \sin(\theta_i)\sin(\phi_i)$, where $i\in\{\mathrm{R}, \mathrm{T}\}$, $k=2\pi / \nu$ is the wave number, $\nu$ the carrier wavelength, whereas $\Delta_\rmx$ and $\Delta_\rmy$ denote the spacing between the antenna elements in the $\rmx$- and $\rmy$-directions, respectively. The steering vector $\a_i(\theta_i,\phi_i) \in \CC^{M_\rmx M_\rmy}$ is given by
\begin{equation*}
    \a_i(\theta_i,\phi_i) = \frac{1}{\sqrt{M_\rmx M_\rmy}}
    \begin{bmatrix} 1 \\ e^{\mathrm{j}\mu^{(i)}_\rmx} \\ \vdots \\ e^{\mathrm{j}(M_\rmx - 1)\mu^{(i)}_\rmx} \end{bmatrix} \kron
    \begin{bmatrix} 1 \\ e^{\mathrm{j}\mu^{(i)}_\rmy} \\ \vdots \\ e^{\mathrm{j}(M_\rmy - 1)\mu^{(i)}_\rmy} \end{bmatrix},
\end{equation*}
where $\kron$ denotes the Kronecker product.
Here, $M_\rmx$ and $M_\rmy$ are the number of antenna elements along the $\rmx$- and $\rmy$-axis. In this paper, we set $\Delta_\rmx = \Delta_\rmy = \nu/2$.

\subsection{Beamforming Codebooks} \label{subsec:codebooks}
We consider analog beamforming with one RF chain at the BS and the UE. 
For simplicity, we employ discrete Fourier transform (DFT) codebooks for precoding and combining.
Let $I_\rmf$ and $I_\rmw$ be the number of precoding and combining beams, while $\bar{\theta}$ and $\bar{\phi}$ are the quantized elevation and azimuth angles, obtained by quantizing the spatial frequencies in the $\rmx$- and $\rmy$-directions such that $2\pi m_\rmx/M_\rmx\le \mu_\rmx \le 2\pi (m_\rmx + M_\rmx - 1)/M_\rmx$, $m_\rmx \in [0,M_\rmx-1]$ and $ 2\pi m_\rmy/M_\rmy \le \mu_\rmy \le 2\pi (m_\rmy + M_\rmy - 1)/M_\rmy$, $m_\rmy \in [0,M_\rmy-1]$, respectively.
Then, the precoding and combining codebooks are given by
\begin{equation}
\begin{aligned}
    \mathcal{C}_{\rm T} &= \Big\{\a_{\rm T}(\bar{\theta}_{\mathrm{T}}^{(1)}, \bar{\phi}_{\mathrm{T}}^{(1)}), \dotsc, \a_{\rm T}(\bar{\theta}_{\mathrm{T}}^{(I_\rmf)}, \bar{\phi}_{\mathrm{T}}^{(I_\rmf)})\Big\} \\
    \mathcal{C}_{\rm R} &= \Big\{\a_{\rm R}(\bar{\theta}_{\mathrm{R}}^{(1)}, \bar{\phi}_{\mathrm{R}}^{(1)}), \dotsc, \a_{\rm R}(\bar{\theta}_{\mathrm{R}}^{(I_\rmw)}, \bar{\phi}_{\mathrm{R}}^{(I_\rmw)})\Big\}, 
    \end{aligned}
\end{equation}
respectively.
%
Let $i_\rmf \in \mathcal{F} = \{1,\dotsc,I_\rmf\}$ and $i_\rmw \in \mathcal{W} = \{1,\dotsc,I_\rmw\}$. 
Given the precoding beam $\f_{i_\rmf} \in \mathcal{C}_{\rm T}$ and the combining beam $\w_{i_\rmw} \in \mathcal{C}_{\rm R}$, the received signal strength (RSS) for the beam pair indexed by $(i_\rmf, i_\rmw)$ is
\begin{equation} \label{eq:rss}
    \mathrm{RSS}_{i_\rmf, i_\rmw} = \Big|\sqrt{P_{\rm T}} \w_{i_\rmw}^\mathsf{H} \H \f_{i_\rmf} s + \w_{i_\rmw}^\mathsf{H} \n \Big|^2.
\end{equation}
In this paper, we set $I_\rmf = M_{\rm T}$ and $I_\rmw = M_{\rm R}$.
\subsection{Dataset Construction}
\begin{figure}
    \centering
    \includegraphics[scale=0.5]{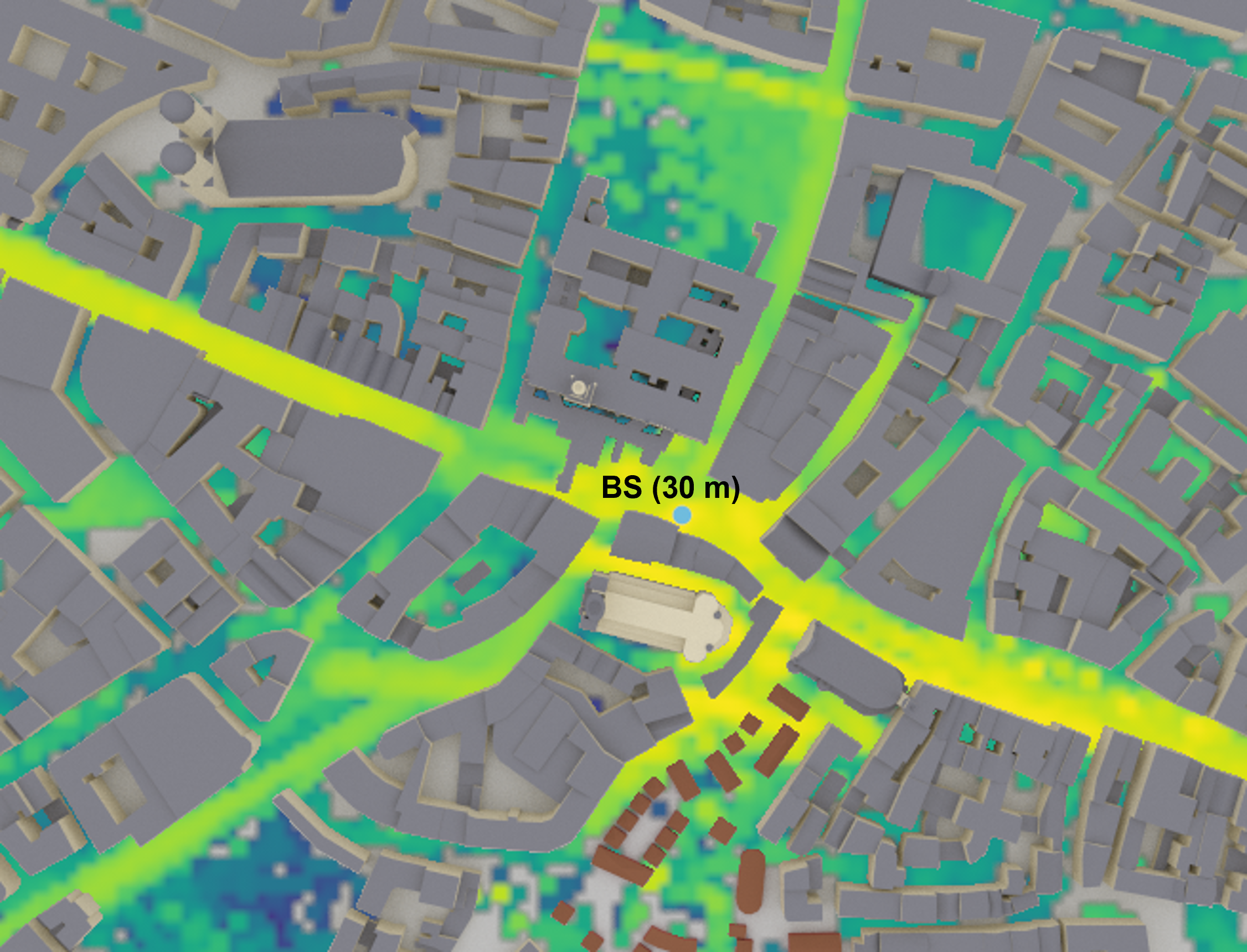}
    \captionsetup{font=small}
    \caption{Illustration of the ray tracing simulation environment  overlaid with a color map showing the coverage of the base station (blue dot). The BS antenna array is placed at a height of 30\,m above the terrain surface. The environment depicts the area around the Frauenkirche in Munich, Germany, and is set up using Sionna \cite{hoydis2022Sionna}.}
    \label{fig:cov_map}
\end{figure}
This paper focuses on an offline learning setting where there is a data collection phase and a training phase.
To construct the dataset, UE positions are randomly sampled within the simulation environment. 
The ray tracing tool simulates multipath propagation to each position and generates the associated MIMO channels.
After generating the channel samples, an exhaustive search is performed to identify the beam pair $(\f_{i^\ast_\rmf}, \w_{i^\ast_\rmw})$ that maximizes the RSS at each UE position.
The optimal beam pair indices $(i^\ast_\rmf, i^\ast_\rmw)$ is thus given by
\begin{equation} \label{eq: exh_search}
    (i^\ast_\rmf, i^\ast_\rmw) = \argmax{i_\rmf \in \mathcal{F}, i_\rmw \in \mathcal{W}} \mathrm{RSS}_{i_\rmf,i_\rmw}.
\end{equation}
The dataset therefore consists of $T$ samples, each containing the position $(x, y, z)$ of the UE and the optimal beam pair indices $(i^\ast_\rmf, i^\ast_\rmw)$.
In practice, UE positions are obtained using Global Positioning System (GPS) sensors which are available in modern user devices. The positions can be shared with the BS using, for example, sub-6 GHz control channels.
\section{Proposed Beam Selection Method} \label{sec:III}
The objective of beam selection in the scenario considered in this paper is to assign the optimal beam pair (or a set of beam pair options) to a UE, given its position.
This may be viewed as a classification problem in which the input features are the UE positions and the output targets (or labels) are the optimal beam pair indices.
ML classifiers based on neural networks attempt to approximate the (possibly highly non-linear) function that maps the features to the labels.
In this paper, we propose to view the UE positions and the beam indices as realizations drawn from an unknown joint probability distribution.
In contrast to non-linear function approximation, we attempt to estimate the underlying joint PMF of the UE positions and the beam indices from the available data samples, after which the estimated PMF can be used for beam selection at new UE positions. \revTwo{We refer to our proposed method as Probabilistic Position-Aided Beam Selection (ProPABS).}  

\subsection{Low-Rank Modeling of the Joint PMF}
Define $\mathsf{X} = \{X_\rmx, X_\rmy, X_\rmf, X_\rmw\}$ consisting of $N=4$ random variables associated with the $\rmx$-coordinate and $\rmy$-coordinate of the UE, the precoding beam index, and the combining beam index, respectively.
Note that in our simulations, the $\mathrm{z}$-coordinate of the UE (i.e., its height from the ground) is constant for all UE positions and can thus be omitted from the collection of random variables.
Let the support of the variables $X_\rmx$ and $X_\rmy$ be uniformly discretized such that $X_\rmx \in [1, I_\rmx]$ and $X_\rmy \in [1, I_\rmy]$.
Furthermore, for ease of notation, define $\mathcal{N} = \{\rmx,\rmy,\rmf,\rmw\}$ as a collection of subscripts.
Then, given that $X_\rmf \in [1, I_\rmf]$ and $X_\rmw \in [1, I_\rmw]$, the joint PMF of $\mathsf{X}$ can be conveniently represented by a four-way tensor $\Xt \in \RR^{I_\rmx \times I_\rmy \times I_\rmf \times I_\rmw}$, where each element of $\Xt$ is the joint probability of a given realization of the four random variables.
In particular, letting $i_n = 1,\dotsc,I_n$ for $n\in \mathcal{N}$, we have that 
%
$\Xt(i_\rmx, i_\rmy, i_\rmf, i_\rmw) = \mathsf{Pr}(X_\rmx=i_\rmx, X_\rmy=i_\rmy,X_\rmf=i_\rmf,X_\rmw=i_\rmw)$.
%

One may choose to estimate $\Xt$ using a histogram.
In this case, the number of parameters to estimate would then be $I_\rmx I_\rmy I_\rmf I_\rmw$, which is at least exponential in $N$ (assuming $I_n > 1,~\forall n$).
Therefore, to reduce the parameter space, we impose a low-rank CPD model (see, e.g., \cite{kolda_tensor_2009}) on $\Xt$.
This decomposition represents $\Xt$ in terms of a sum of rank-one tensors, i.e.,
\begin{equation} \label{eq:CPD}
    \Xt = \sum_{r=1}^R \lambda_r \A_\rmx(:,r)\circ\A_\rmy(:,r)\circ\A_\rmf(:,r)\circ\A_\rmw(:,r),
\end{equation}
where $\circ$ denotes the outer product, $R$ is the smallest number of rank-one components for which such a decomposition exists, $\lambdab=[\lambda_1,\dotsc,\lambda_R]^\mathsf{T}$ is a \textit{loading vector} whose elements $\lambda_r$ scale each rank-one component, while $\A_n(:,r)$ is the $r$-th column of the $n$-th \textit{factor matrix} $\A_n \in \RR^{I_n \times R}$.
A particular element of $\Xt$ is given by
\begin{equation}
    \Xt(i_x,i_y,i_\rmf,i_\rmw) = \sum_{r=1}^R \lambda_r \prod_{n \in \mathcal{N}} \A_n(i_n,r).
\end{equation}

Kargas \textit{et al.} have shown in \cite{kargas_tensors_2018} that such a decomposition of a PMF tensor can be interpreted as a na\"{i}ve Bayes model with one latent variable $H$ which takes $R$ states.
Under this interpretation, $\lambda_r = \mathsf{Pr}(H=r)$, i.e, the prior probability that the latent variable takes the $r$-th state, while $\A_n(:,r) = \mathrm{p}(X_n\,|\,H=r)$, i.e., the conditional PMF of the $n$-th variable, given that $H=r$.
Thus, the decomposition in \eqref{eq:CPD} is subject to nonnegativity ($\lambdab > \bm{0}, \A_n \ge \bm{0},\,\forall n$) and sum-to-one ($\bm{1}^\mathsf{T} \lambdab = 1, \bm{1}^\mathsf{T} \A_n = \bm{1}^\mathsf{T},\,\forall n$) constraints (also referred to as probability simplex constraints).
The total number of free parameters in the CPD representation is $(R-1) + \sum_{n \in \mathcal{N}} R(I_n - 1)$, which is linear in $N$ rather than exponential, as is the case with histogram estimation.
A distinctive feature of such a nonnegative CPD is that it is essentially unique up to a permutation ambiguity among the rank-one components (see, e.g., \cite{kolda_tensor_2009}).

\subsection{Estimating the Joint PMF} 
In this subsection, we describe how to estimate the model parameters $\{\lambdab, \A_\rmx, \A_\rmy, \A_\rmf, \A_\rmw\}$ from training data using Bayesian inference \cite{chege_bayesian_2023}.
%
%
Define the dataset consisting of $T$ i.i.d. samples of the UE positions $(x,y)$ and the optimal beam indices $(i^\ast_\rmf,i^\ast_\rmw)$ as $\mathcal{D} = \{\d_t\}_{t=1}^T$, where $\d_t = [x_t,y_t, i^\ast_{\rmf,t}, i^\ast_{\rmw,t}]^\mathsf{T}$ and $(\cdot)^\mathsf{T}$ denotes the transpose operator. 
%
%
%
Under the na\"{i}ve Bayes interpretation, each sample $\d_t$ is associated with a realization of the latent variable $H\in [1,R]$. 
We define a \textit{local} latent variable $\z_t=[z_{1,t},\dotsc,z_{R,t}]^\mathsf{T}$ such that $z_{r,t}=1$ if $H=r$ and $z_{r,t}=0$ otherwise. It follows that $\mathsf{Pr}(z_{r,t}=1) = \mathsf{Pr}(H=r)=\lambda_r$. 

From the definition of $\z_t$, we can write $\mathrm{p}(\z_t\,|\,\lambdab) = \prod_{r=1}^R\lambda_r^{z_{r,t}}$.
The likelihood of $\mathcal{D}$ given the parameters $\THETA = \{\Z, \lambdab, \A_\rmx, \A_\rmy, \A_\rmf, \A_\rmw\}$ is
\begin{equation}
    \mathrm{p}(\mathcal{D}\,|\,\THETA) = \mathrm{p}(\mathcal{D}\,|\,\Z, \A_\rmx, \A_\rmy, \A_\rmf, \A_\rmw) \cdot \mathrm{p}(\Z\,|\,\lambdab),
\end{equation}
where $\Z=\{\z_t\}_{t=1}^T$, $\mathrm{p}(\Z\,|\,\lambdab)  = \prod_{t=1}^T \mathrm{p}(\z_t\,|\,\lambdab)$ and
%
\begin{equation*}
        \begin{aligned}
         \mathrm{p}(\mathcal{D}\,|\,\Z, \A_\rmx, \A_\rmy, \A_\rmf, \A_\rmw) & = \\
        & \hspace{-10em} \prod_{t=1}^T \prod_{r=1}^R \Big(\A_\rmx(x_t,r) \A_\rmy(y_t,r) \A_\rmf(i^\ast_{\rmf,t},r) \A_\rmw(i^\ast_{\rmw,t},r)\Big)^{z_{r,t}}. 
        \end{aligned}
\end{equation*}
%
To ensure that the probability simplex constraints are satisfied, we assign Dirichlet prior distributions (see, e.g., \cite{bishop2006}) for $\lambdab$ and $\A_n(:,r), \forall n,r$, i.e., $$\mathrm{p}(\lambdab) \propto \prod_{r=1}^R \lambda_r ^ {\alpha_{\lambda,r}}, ~ \mathrm{p}\Big(\A_n(:,r)\Big) \propto \prod_{i_n=1}^{I_n} \Big(\A_n(i_n,r)\Big)^{\alpha_{n,r,i_n}},$$ where $\alpha_{\lambda,r}$ and $\alpha_{n,r,i_n}$ are the hyperparameters of the priors. 

The posterior distribution of the model parameters given the data is found using Bayes' theorem, $\mathrm{p}(\THETA\,|\,\mathcal{D}) = \mathrm{p}(\mathcal{D},\THETA)/\int_{\THETA}\mathrm{p}(\mathcal{D},\THETA)$, where 
\begin{equation}
\mathrm{p}(\mathcal{D},\THETA)=\mathrm{p}(\mathcal{D}\,|\,\THETA) \cdot \mathrm{p}(\lambdab) \cdot \prod_{n \in \sysN}\prod_{r=1}^R\mathrm{p}\Big(\A_n(:,r)\Big).
\end{equation}
Exact Bayesian inference is infeasible due to the high-dimensional integral in the denominator of Bayes' theorem. 
Therefore, in this work, we adopt the variational Bayes for PMF estimation (VB-PMF) algorithm proposed in \cite{chege_bayesian_2023} to approximate $\mathrm{p}(\THETA\,|\,\mathcal{D})$ using variational inference techniques. 
Finally, point estimates $\{\hat{\lambdab}, \hat{\A}_\rmx, \hat{\A}_\rmy, \hat{\A}_\rmf, \hat{\A}_\rmw\}$ are obtained by computing conditional expectations with respect to the estimated posterior distributions for each parameter.

An important advantage of the VB-PMF algorithm is that the rank $R$ of $\Xt$, which is usually unknown in practical scenarios, is estimated as part of the inference procedure.
In particular, thanks to the sparsity-promoting property of the Dirichlet distribution, setting $\alpha_{\lambda,r} < 1, \forall r$ ensures that the estimated posterior distribution for $\lambdab$ (i.e., the PMF of $H$) is sparse. We choose $\alpha_{\lambda,r} = 10^{-6}, \forall r$.
On the other hand, we set $\alpha_{n,r,i_n}=1, \forall n,r,i_n$ to allow the columns $\A_n(:,r)$ (i.e., the conditional PMFs) to explore the space of all possible distributions.
The algorithm is initialized with a rank greater than the unknown true rank (e.g., the maximum possible rank required for CPD uniqueness).
Since the joint PMF tensor is low-rank, some elements of $\hat{\lambdab}$ will be close to zero after convergence.
The corresponding rank-one components can then be removed, resulting in automatic rank detection.
Due to space limitations, we refer the interested reader to \cite{chege_bayesian_2023} for more details.
\subsection{PMF-Based Beam Selection}
Given a test UE position $(x_{\rm test},y_{\rm test})$, we would like to predict the optimal precoding and combining beam indices $(\hat{i}^\ast_\rmf, \hat{i}^\ast_\rmw)$. 
The parameters $\{\hat{\lambdab}, \hat{\A}_\rmx, \hat{\A}_\rmy, \hat{\A}_\rmf, \hat{\A}_\rmw\}$ represent an estimate of the joint PMF of all UE positions, precoding beam indices, and combining beam indices.  
With the joint PMF at hand, we can obtain $(\hat{i}^\ast_\rmf, \hat{i}^\ast_\rmw)$ by maximizing the posterior distributions of the beam indices given the test position.
For a classification problem, such an estimate, referred to as the maximum \textit{a posteriori} (MAP) estimate, minimizes the probability of misclassification \cite{bishop2006}.
Thus, we have
\begin{subequations} \label{eq:map}
    \begin{align}
        (\hat{i}^\ast_\rmf, \hat{i}^\ast_\rmw) & = \argmax{i_\rmf \in \sysF,i_\rmw \in \mathcal{W}} \mathsf{Pr}(i_\rmf, i_\rmw\,|\, x_{\rm test}, y_{\rm test}) \label{eq:map_a}  \\
        & \hspace{-3em}= \argmax{i_\rmf \in \sysF,i_\rmw \in \mathcal{W}} \sum_{r=1}^R \mathsf{Pr}(H=r) \mathsf{Pr}(x_{\rm test}, y_{\rm test}, i_\rmf, i_\rmw\,|\,r) \label{eq:map_b} ,
    \end{align}
\end{subequations}
where \eqref{eq:map_b} is obtained after applying Bayes' rule and noting that the maximization is independent of the marginal probability $\mathsf{Pr}(x_{\rm test}, y_{\rm test})$.
%
Recall that under the na\"{i}ve Bayes model, the variables $\{X_\rmx,X_\rmy,X_\rmf,X_\rmw\}$ are independent when conditioned on the latent variable $H$. 
Thus, the objective in~\eqref{eq:map_b} is expressed in terms of the estimated model parameters as
\begin{equation}
    \sum_{r=1}^R \hat{\lambda}_r \hat{\A}_\rmf(i_\rmf,r) \hat{\A}_\rmw(i_\rmw,r) \hat{\A}_\rmx(x_{\rm test},r) \hat{\A}_\rmy(y_{\rm test},r).
\end{equation}
The estimated posterior distribution $\mathrm{p}(X_f, X_w\,|\, x_{\rm test}, y_{\rm test})$ is used to produce a top-$N_{\rm b}$ beam pair list by selecting the indices corresponding to the $N_{\rm b}$ largest posterior probabilities.

\section{Results and Discussion} \label{sec:IV}

We consider the simulated outdoor environment depicted in Fig.~\ref{fig:cov_map}.
The transmit power $P_{\rm T}$, the carrier frequency $f_c$, and the carrier bandwidth $B$ are set to 30\,dBm, 26\,GHz, and 200\,MHz, respectively.
Furthermore, we assume that the noise variance is given by $\sigma_n^2 = -174 + 10\log_{10} B$ dBm.
The BS and the UE are equipped with $8 \times 8$ and $2 \times 2$ UPAs, respectively.
Thus, the number of antennas is $M_{\rm T} = 64$ and $M_{\rm R} = 4$.
The positions of the UEs are randomly sampled within the BS coverage area and realistic channel responses between the BS and each position are generated using the Sionna ray tracing module \cite{hoydis2022Sionna}.
For each UE position, the RSS is calculated using all precoder-combiner combinations according to \eqref{eq:rss}. 
The optimal beam pair indices are then obtained using \eqref{eq: exh_search}.

The dataset $\mathcal{D}$ consists of $2\times 10^3$ samples, of which 80\,\% are used for training and 20\,\% for testing.
Moreover, \revTwo{50} experiments are conducted, each involving the shuffling of the entire dataset and its division into training and testing sets, with the final result being the average across all experiments.
%
%
For \revTwo{ProPABS}, the $\rmx$- and $\rmy$-coordinates are discretized \revTwo{into bins of size 5\,m, yielding $I_\rmx=279$ and $I_\rmy=165$}, respectively.
%
%
We initialize the VB-PMF algorithm with $R=30$ components. \revTwo{The estimated ranks obtained from 50 experiments, along with their frequency of occurrence, are $\hat{R}=4~(22\%)$, $\hat{R}=5~(62\%)$, and $\hat{R}=6~(16\%)$.}
%
%
Then, we compare \revTwo{ProPABS} with two position-aided beam selection approaches: 
\begin{enumerate}
    \item Inverse multipath fingerprinting \cite{va_inverse_2018}, which builds a database of the top beam pair indices for each UE position after performing an exhaustive search as in \eqref{eq: exh_search}. During training, the beam pairs are ranked according to their probability of being optimal (i.e., having the highest RSS) in the database.  The ranked list of candidate beam pairs is then considered for selection during the testing phase. 
    \revTwo{Similar to the setting in our method, a bin size of 5\,m is used to discretize the UE positions.} 

    \item A fully connected neural network (NN), where the UE positions are input features while the optimal beam pairs are output labels (e.g., \cite{heng_machine_2021}). 
    We employ an NN with three hidden layers consisting of 6, 18, and 48 neurons, respectively, with sigmoid activation functions. The input layer has 3 neurons corresponding to the UE position, while the output layer has $M_{\rm T}M_{\rm R} = 256$ neurons. The NN is trained using the Adam algorithm~\cite{DBLP:journals/corr/KingmaB14}.   
\end{enumerate}

Each beam selection method produces a candidate list $\mathcal{S}$ of the top-$N_{\rm b}$ beam pairs and the pair resulting in the highest RSS is selected, i.e.
\begin{equation}
    (\hat{i}^\ast_\rmf, \hat{i}^\ast_\rmw) = \argmax{(\hat{i}_\rmf, \hat{i}_\rmw) \in \mathcal{S}} \mathrm{RSS}_{\hat{i}_\rmf, \hat{i}_\rmw}.
\end{equation}
To evaluate the performance of the beam selection methods, we consider the power loss probability, defined as follows.
Let the set of all possible beam pairs be $\mathcal{B}$.
Then, the power loss probability is given by \cite{va_inverse_2018}
\begin{equation}
    P_{\rm pl}(c, \mathcal{S}) = \mathsf{Pr}\Big[\max{(i_\rmf, i_\rmw) \in \mathcal{B}} \mathrm{RSS}_{i_\rmf,i_\rmw} > c\cdot\mathrm{RSS}_{\hat{i}^\ast_\rmf, \hat{i}^\ast_\rmw}\Big]
\end{equation}
where $c \ge 1$.
In our experiments, we evaluate the 0\,dB ($c=1$) and 3\,dB ($c=2$) power loss probabilities. 
In addition, we compute the achievable rate of the selected beam pair, defined as
\begin{equation}
    R_{\rm a} = \log_2\Big[1 + \Big(\big|\sqrt{P_{\rm T}}\w_{\hat{i}^\ast_\rmw}^\mathsf{H}\H\f_{\hat{i}^\ast_\rmf} s\big|^2\big/\sigma_n^2\Big)\Big].
\end{equation}

\begin{figure*}[t]
    \centering
    \begin{subfigure}{0.32\textwidth}
        \centering
        \includegraphics[scale=0.2]{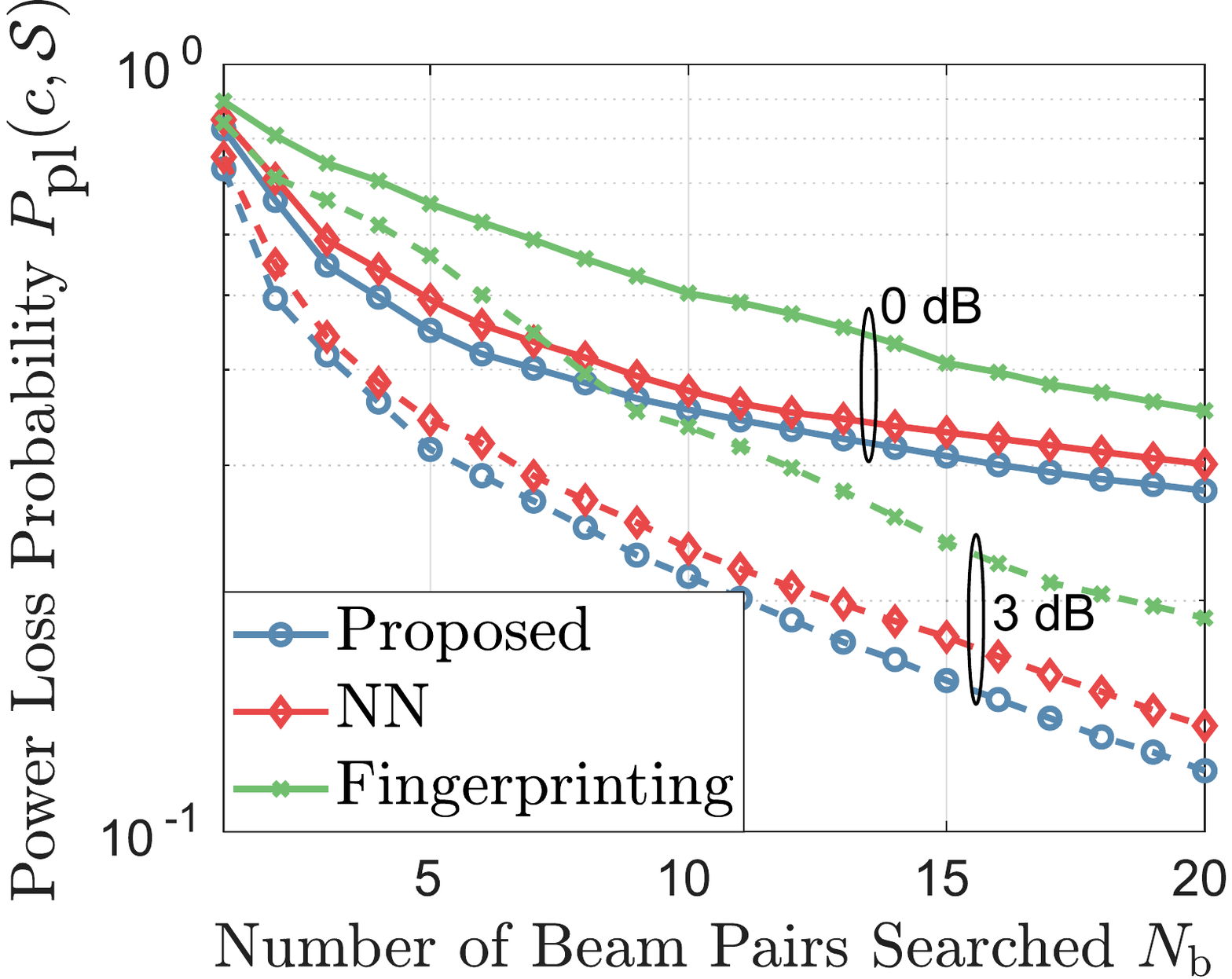}
        \caption{}  
        \label{fig:pl_vs_Nb}
    \end{subfigure}
    \hfill
    \begin{subfigure}{0.32\textwidth}
        \centering
        \includegraphics[scale=0.205]{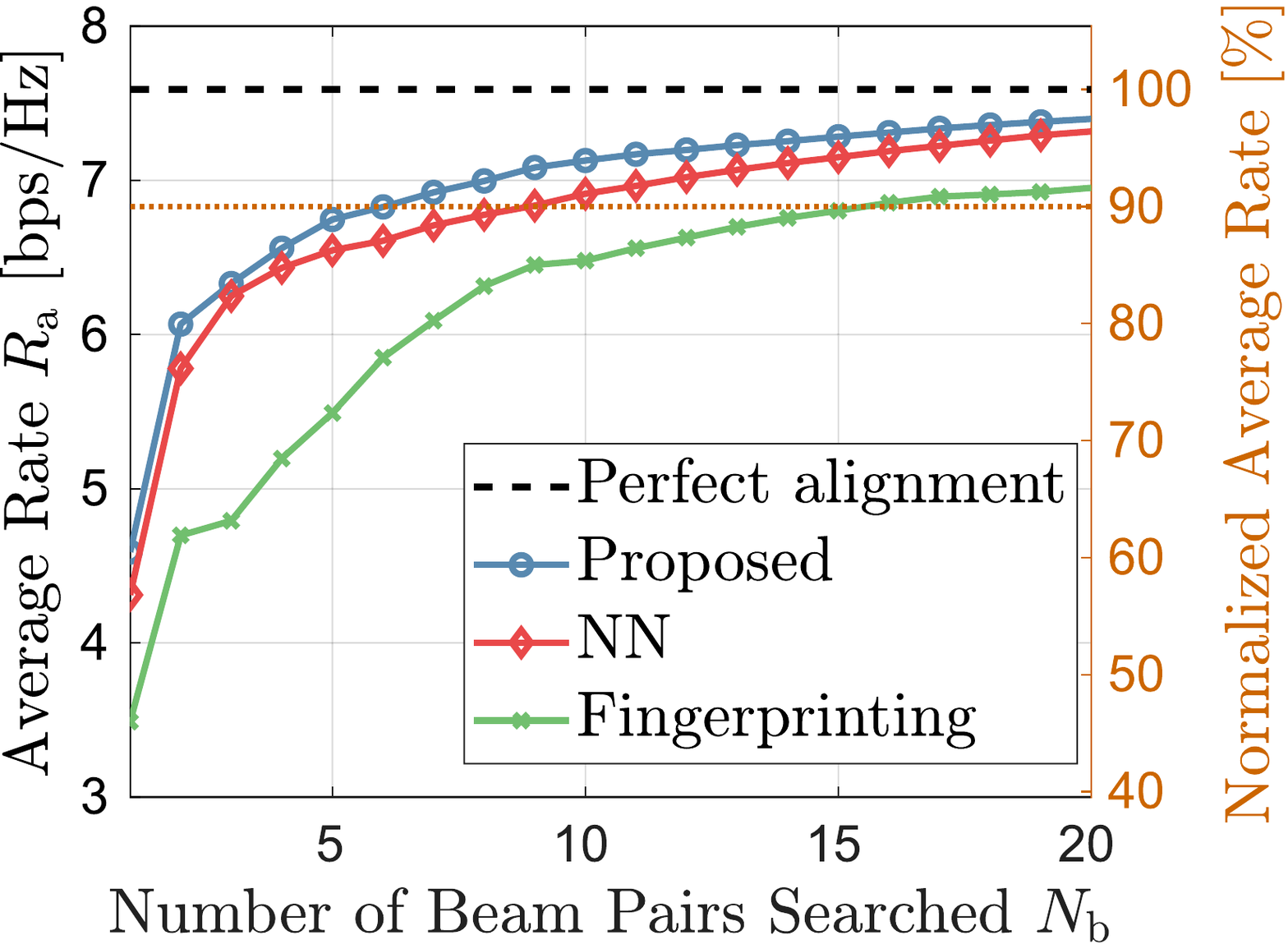}
        \vspace{-1.6em}
        \caption{}
        \label{fig:rate_vs_Nb}
    \end{subfigure}
    \hfill
    \begin{subfigure}{0.32\textwidth}
        \centering
        \includegraphics[scale=0.2]{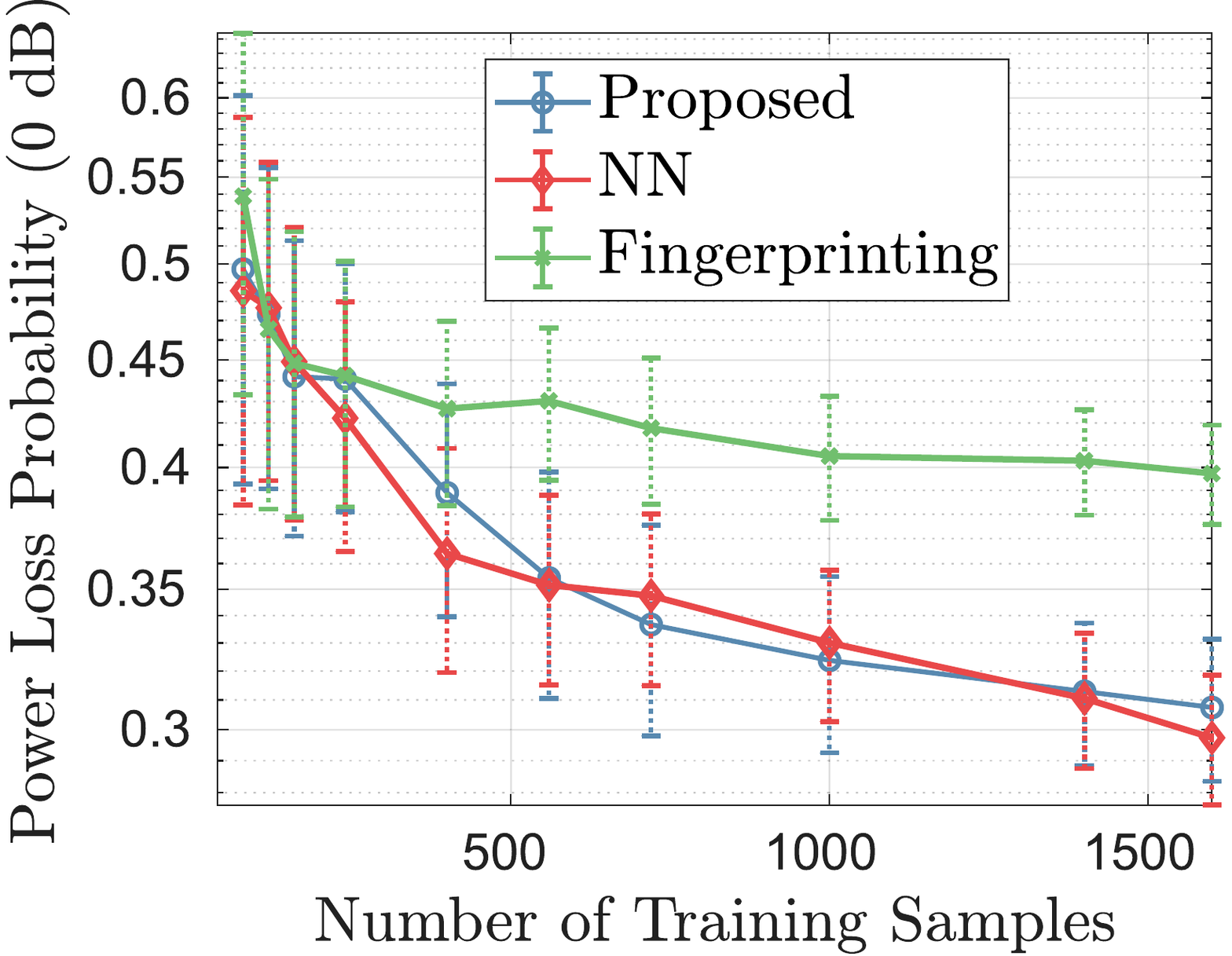}
        \caption{}
        \label{fig:pl_vs_train}
    \end{subfigure}
    \vspace{0.5em}
    \captionsetup{font=small}
    \caption{Performance evaluation of the proposed beam selection method compared to two baselines.}
    \label{fig:2}
    \vspace{-0.5em}
\end{figure*}

Fig.\,\ref{fig:2} presents the results arising from our numerical simulations.
Fig.\,\ref{fig:pl_vs_Nb} shows the average power loss probability as a function of the number of beam pairs searched.
\revTwo{ProPABS} consistently achieves a lower power loss probability compared to the baseline approaches.
Fig.\,\ref{fig:rate_vs_Nb} presents the average achievable rate and the corresponding average rate normalized by the perfect beam alignment case.
We observe that \revTwo{ProPABS} already achieves about 90\,\% of the maximum achievable rate by searching only the top $N_{\rm b} = 6$ beam pairs.
Note that for the setup considered, there are $M_{\rm T}M_{\rm R} = 256$ possible beam pair combinations, demonstrating that \revTwo{ProPABS} greatly reduces the search space while achieving a considerably high average rate.
In comparison, the NN and the fingerprinting method need to search around $N_{\rm b} = 9$ and $N_{\rm b} = 16$ beam pairs, respectively, to achieve a similar normalized average rate. 
Fig.\,\ref{fig:pl_vs_train} shows the average 0\,dB power loss probability as a function of the number of training samples, ranging from 80 to 1600.
For this evaluation, we set $N_{\rm b}=16$ to ensure a normalized average rate greater than 90\,\% for all beam selection methods \revTwo{ and average over 1000 trials}.
We observe a marked improvement for \revTwo{ProPABS} up to around 600 training samples, followed by a gradual improvement afterwards.
\revTwo{The NN performs similarly to our method,}
while the fingerprinting method would need a much larger candidate beam list to benefit from more training samples.
\revTwo{Therefore, \revTwo{ProPABS} can achieve an acceptable average rate with relatively few training samples, demonstrating its efficiency.}
%

\section{Conclusion} \label{sec:V}
This paper presents an efficient beam selection method for mmWave MIMO systems that employs a low-rank probability mass function (PMF) tensor model \revTwo{with interpretable parameters}.
By estimating the joint PMF tensor of discretized UE positions and beam indices using a Bayesian approach, the proposed method effectively reduces the beam search space while maintaining a high achievable rate.
Compared to neural network and fingerprinting-based approaches, our method achieves \revTwo{a better performance in terms of the power loss probability and the achievable rate} with \revTwo{relatively few training} samples.
The effectiveness of the proposed approach is demonstrated using data generated from a state-of-the-art ray tracing simulator, ensuring realistic evaluation conditions.
%
%
A possible direction for future work is to extend the approach to online beam selection, allowing continuous adaptation to the communication environment.  

\bibliographystyle{IEEEtran}
\bibliography{refs.bib}

\end{document}